\documentclass[pra,twocolumn,showpacs]{revtex4}

\usepackage{amsmath,amssymb,mathrsfs}
\usepackage{graphicx}
\usepackage{amsmath}
\usepackage{epsfig}
\usepackage[usenames]{color}
\usepackage{ulem}
\usepackage[colorlinks=true,citecolor=blue]{hyperref}

\newcommand{\beq}{\begin{equation}}
\newcommand{\eeq}{\end{equation}}
\newcommand{\beqa}{\begin{eqnarray}}
\newcommand{\eeqa}{\end{eqnarray}}
\newcommand{\ba}{\begin{array}}

\newcommand{\ea}{\end{array}}

\renewcommand{\b}[1]{\mathbf{ #1}}

\newcommand{\cop}[1]{{#1^{\dagger}}\!}  						
\newcommand{\aop}[1]{{#1}}

\begin{document}

\title{Tunable zero and first sounds in ultracold Fermi gases 
with Rabi coupling}

\author{L. Lepori}
\affiliation{Dipartimento di Fisica e Astronomia ``Galileo Galilei'' 
and CNISM, Universit\`a di Padova, Via Marzolo 8, 35131 Padova, Italy,}
\affiliation{Dipartimento di Scienze Fisiche e Chimiche, Universit\`a dell'Aquila, via Vetoio,
I-67010 Coppito-L'Aquila, Italy}
\affiliation{INFN, Laboratori Nazionali del Gran Sasso, Via G. Acitelli, 22, I-67100 Assergi (AQ), Italy.}
 
\author{L. Salasnich}
\affiliation{Dipartimento di Fisica e Astronomia ``Galileo Galilei'' 
and CNISM, Universit\`a di Padova, Via Marzolo 8, 35131 Padova, Italy}
\affiliation{CNR-INO, via Nello Carrara, 1 - 50019 Sesto Fiorentino, Italy.}

\begin{abstract}
We consider a weakly-interacting fermionic gas of alkali-metal atoms 
characterized by two hyperfine states which are Rabi coupled. 
By using a Hartree approximation for the repulsive interaction  
we determine the zero-temperature equation of state of this 
Fermi gas in $D$ spatial dimensions ($D=1,2,3$). Then, 
adopting the Landau-Vlasov equation and 
hydrodynamic equations, we investigate the speed of first sound and zero sound. 
We show that the two sounds, which occur respectively in 
collisional and collisionless regimes, crucially depend on the 
interplay between interaction strength and Rabi coupling. 
Finally, we discuss for some experimentally relevant cases 
the effect of a trapping harmonic potential on the density profiles 
of the fermionic system. 
\end{abstract}


\maketitle

\section{Introduction}

In the last few years the study of ultracold alkali-metal atoms was made even 
more stimulating by the experimental advent of synthetic gauge potentials, 
applied on multi-component gases in different hyperfine levels 
\cite{spielman2011,spielman2013,dalibard2011,lewensteinbook}. 
A related ingredient, experimentally quite simple to implement 
because it does not involves space-dependent tunnelling processes, 
is the Rabi coupling. This technique is nowadays a common tool 
for experimental and theoretical investigations involving 
multi-component gases. Some examples are 
the control of the population of the hyperfine levels 
\cite{steck}, the formation of localized structures 
\cite{horstman2010}, and the mixing-demixing dynamics of 
Bose-Einstein condensates \cite{ober,abad2013}. 
 
Due to the wide applicability of the Rabi coupling, it is 
particularly interesting to investigate how the presence of a Rabi 
term affects equilibrium and collective dynamical properties 
of interacting atomic gases.
In the present paper we analyze the effect 
of a Rabi coupling on a two-hyperfine-component Fermi gas.  
In particular we focus on zero suond and first sound \cite{landaustat}, 
which can be experimentally obtained with a local perturbation 
of the gas density. In general, the study of sound propagation 
in Fermi liquids, from Helium to electrons to cold gases, 
is of extreme importance to understand the physical properties 
of the system \cite{landaustat,leggett}. 

We determine at first the equation of state for the gas, 
in the presence of a weak repulsive interaction within 
a Hartree approximation. We find that the Rabi coupling term 
divides the spectrum in two branches with 
different energies. Later on, we use this equation of state 
to derive the behavior of the zero and first sounds. 
In this analysis we adopt hydrodynamic equations in the 
collisional regime and the Landau-Vlasov 
equation in the collisionless regime \cite{landaustat,leggett}. 
Finally, in the Appendix, density profiles 
in presence of an harmonic trap, mostly used in current experiments, 
are derived within the local density approximation, as well as the densities 
and the chemical potentials at the center of the trap, 
where the measurements for the sounds are usually performed.

\section{The model}
\label{model}

We consider a two-spin-component Fermi gas of alkali-metal atoms 
with mass $m$, described by the Hamiltonian 
\begin{multline}
{\hat H} =\int \mathrm{d}\b{r}  \Bigg\{ \sum_{\sigma = \uparrow, \downarrow} 
\cop{{\hat \psi}}_{\sigma} (\b{r})  \, \Big(- \frac{   \hbar^2 }{2m} \, 
\nabla^2 
- \mu  \Big) \, \aop{{\hat \psi}}_{\sigma}  (\b{r})  
\\
+  \hbar \Omega \,  \Big( \cop{{\hat \psi}}_{\uparrow}  (\b{r}) 
\aop{{\hat \psi}}_{\downarrow} 
(\b{r}) + \cop{{\hat \psi}}_{\downarrow}  (\b{r}) 
\aop{{\hat \psi}}_{\uparrow}(\b{r}) 
\Big) + g  \, \aop{{\hat n}}_{\uparrow} (\b{r}) \,  
\aop{{\hat n}}_{\downarrow} (\b{r}) 
\Bigg\} \, ,
\label{H_cont}
\end{multline}
where ${\hat \psi}_{\sigma}({\bf r})$ is the fermionic field operator 
for atoms with spin $\sigma$, 
${\hat n}_{\sigma} (\b{r}) = {\hat \psi}^{\dagger}_{\sigma}(\b{r}) 
{\hat \psi}_{\sigma}(\b{r})$ is the local number density operator, 
and $\mu$ is the chemical potential \cite{lewensteinbook}. 
The term proportional to $\hbar \Omega$ corresponds to the Rabi 
coupling inducing a spin flip between the components, while the term 
proportional to $g$ models the density-density inter-component 
repulsion ($g>0$). 

We assume that system is in a $D$-dimensional space. 
Moreover, we work in the weak-coupling regime $gn \ll 1$, with 
$n=\langle {\hat n}_{\uparrow}(\b{r})\rangle+ 
\langle {\hat n}_{\downarrow}(\b{r})\rangle$ 
the average total number density, to avoid Stoner 
instability \cite{stoner} and itinerant ferromagnetism \cite{itinero1}. 
In this way, at equilibrium, we can safely set 
$\langle {\hat n}_{\uparrow}(\b{r})\rangle = 
\langle {\hat n}_{\downarrow}(\b{r})\rangle$.  

\section{Equation of state}
\label{equil}

\subsection{Non-interacting case}
\label{free}

Let us start with the very simple case of non-interacting fermions. 
If $g = 0$ the Hamiltonian in Eq. (\ref{H_cont}) is quadratic, then 
it can be diagonalized in momentum space by a global 
unitary transformation, obtaining 
\beq
{\hat H} =\sum_{s = \pm} \, \sum_{\bf k}  \lambda_{s}(\b{k})  
\, {\hat \eta}^{\dagger}_s (\b{k}) {\hat \eta}_s (\b{k})
\label{H_cont2}
\eeq
with
\beq
\lambda_{\pm}(\b{k}) = \frac{\hbar^2 |\b{k}|^2}{2m} - \big(\mu \pm \hbar 
\, \Omega\big) \, .
\label{spec}
\eeq
We  have in particular that ${\hat \eta}_{\pm}
(\b{k}) = ({\hat \psi}_{\uparrow}(\b{k}) 
\mp {\hat \psi}_{\downarrow}(\b{k}))/\sqrt{2}$ and the application of the Rabi 
coupling leads to an effective unbalance equal to $2 \, \Omega$ in the 
rotated components $\eta_{\pm}$, a trivial result for people working on 
quantum optics. More in detail, in the present case
we see from Eq. (\ref{spec}) that the spectrum divides in two branches 
with different energies.  However when $\mu < \hbar \Omega$ only the 
eigenstates with energies $\lambda_{+}(\b{k})$ are occupied.

Equivalently, we obtain two free fermionic gases with the same energies 
$ \frac{\hbar^2 |\b{k}|^2}{2m}$ and different 
chemical potentials $\mu_{\pm} =  \big(\mu \pm \hbar \, \Omega \big)$. 
 
As for a free one component Fermi gas,  in this case the effective chemical 
potentials $\mu_{\pm}$ are linked to the zero-temperature average 
densities $n_{\pm}$ by the equations
\beq
\mu_{\pm} = \big(\mu \pm \hbar \, \Omega\big) = 
\Big(\frac{n_{\pm}}{A_D} \Big)^{\frac{2}{D}} \, ,
\label{mudens}
\eeq
where the constant $A_D$ depends on the space dimension $D$. 
In particular, one has 
$A_1 = \frac{\sqrt{2} \, m^{\frac{1}{2}}}{\pi \hbar } $, $A_2 = 
\frac{m}{2 \pi \hbar^2} $ and $A_3 = \frac{\sqrt{2}}{3 \pi^2 \hbar^3} 
\, m^{\frac{3}{2}}$. 

These equations fix the chemical potential $\mu$, 
via the related equations
\beq
\frac{n}{A_D} = \big(\mu + \hbar \Omega \big)^{\frac{D}{2}} + 
\big(\mu - \hbar \Omega \big)^{\frac{D}{2}}  \, .
\label{eqmu2}
\eeq
When $\mu < \hbar \Omega$, only the eigenstates with energies 
$\lambda_{+}(\b{k})$ are occupied, then Eqs. (\ref{eqmu2}) reduce to 
\beq
\frac{n}{A_D} = \big(\mu + \hbar \Omega \big)^{\frac{D}{2}}  \, .
\label{eqmu2sing}
\eeq
It is useful to rescale the energies $\mu$ and $\hbar \Omega$  
by an energy scale $\Lambda_D$, introducing in Eqs. (\ref{eqmu2}) 
and (\ref{eqmu2sing}) the adimensional quantities
$\tilde{\mu} \equiv \frac{\mu}{\Lambda_D}$, $\tilde{n} \equiv 
\frac{n}{A_D \, \Lambda_D^{\frac{D}{2}}}$,
$\tilde{\Omega} \equiv \frac{\hbar \Omega}{\Lambda_D}$. 
The energy scale  $\Lambda_D$ is defined finally as $\Lambda_D 
= \frac{\hbar^2}{2 m} \,  n^{\frac{2}{D}}$.\\
Equations (\ref{eqmu2}),
rescaled by $\Lambda_D$, read:
\beq
\tilde{n}= \big(\tilde{\mu} + \tilde{\Omega}\big)^{\frac{D}{2}} 
+ \big(\tilde{\mu} - \tilde{\Omega}\big)^{\frac{D}{2}} \equiv 
\tilde{\mu}_{+} ^{\frac{D}{2}} + \tilde{\mu}_-^{\frac{D}{2}}\, .
\label{eqxtilde}
\eeq
In the particular case $D= 2$ we obtain immediately $n = 2 
\tilde{\mu}$, not depending on $\Omega$.

\subsection{Interacting case}
\label{freeint}

If $g \neq 0$  the {\it global} unitary transformation 
diagonalizing the quadratic part of Eq. (\ref{H_cont}) maps the 
interaction term as follows:
\beq
g \int \mathrm{d} \b{r} \, \aop{{\hat n}}_{\uparrow} (\b{r}) \,  
\aop{{\hat n}}_{\downarrow} 
(\b{r}) \to g \int \mathrm{d} \b{r} \, \aop{{\hat n}}_{+} (\b{r}) \,  
\aop{{\hat n}}_{-} (\b{r}) \, .
\label{rotint}
\eeq
We see that the interaction term transforms covariantly, as a 
consequence of the fermionic nature of field 
operators and of the global nature of the unitary transformation 
leading to Eq. \eqref{H_cont2}. The latter features occurs only 
because  the Rabi term proportional to $\Omega$ 
in Eq. (\ref{H_cont}) is the same in 
every point of the space. This is not the case for instance in 
the presence of a spin-orbit coupling.

We simplify the repulsive interaction term  keeping only the Hartree terms in the first Born approximation:
$g \, {\hat n}_{+}({\bf r}) \, {\hat n}_{-}({\bf r}) \simeq 
g \, {\hat n}_{+}({\bf r}) \, n_{-} + g \, {\hat n}_{-}({\bf r}) \, n_{+}$, 
where $n_{\pm}=\langle{\hat n}_{\pm}({\bf r}) \rangle$. 
In this way the chemical potentials $\mu_{\pm}$ of the 
interacting system become
\beq
\mu_{\pm} \to \mu_{\pm} - g \,    n_{\mp} =  \mu_{\pm} - g  \,  
\big( n - n_{\pm} \big)  \equiv \mu^{(I)}_{\pm}\, .
\label{shift}
\eeq
The label $(I)$, denoting the chemical potentials shifted by the 
effect of the interaction, will be also used in the following 
for the potentials  $\mu_{\pm}^{(I)}$.
The (dependent) quantities $\mu$, $\mu_{\pm}$ and  $\mu^{(I)}_{\pm}$, 
as well as the densities $n_{\pm}$, can be 
obtained solving two equations, similar to Eq. \eqref{mudens}:
\beq 
n_{\pm} = A_D \, \Big(\mu \pm  \hbar \, \Omega  - g \, (n-n_{\pm}) 
\Big)^{\frac{D}{2}}  \, ,
\label{eqmuintot}
\eeq
or, in adimensional form
($\tilde{n}_- = \tilde{n} - \tilde{n}_+ $):
\beq 
\tilde{n}_{\pm} = \Big(\tilde{\mu} \pm  \tilde{\Omega}  - \tilde{g} 
\, (\tilde{n}-\tilde{n}_{\pm}) \Big)^{\frac{D}{2}}  \, ,
\label{eqmuintsbil2}
\eeq
where $\tilde{g} \equiv g \Big(A_D/\Lambda_D^{1-\frac{D}{2}}\Big)$. 

If $D=2$ Eqs. (\ref{eqmuintot}) can be solved also analitically, 
yielding the results 
\beq
n_{\pm} = \frac{n}{2} \pm \frac{\hbar \, \Omega \, A_2}{1 - g A_2}
\eeq
and 
\beq
\mu = \frac{n}{2 \, A_2} \, (1+ g \, A_2) \,  .
\label{mu2d}
\eeq
No dependence on $\Omega$ is found for $\mu$ in this case. 

\begin{figure}[!ht]
\includegraphics[width=0.47\textwidth]{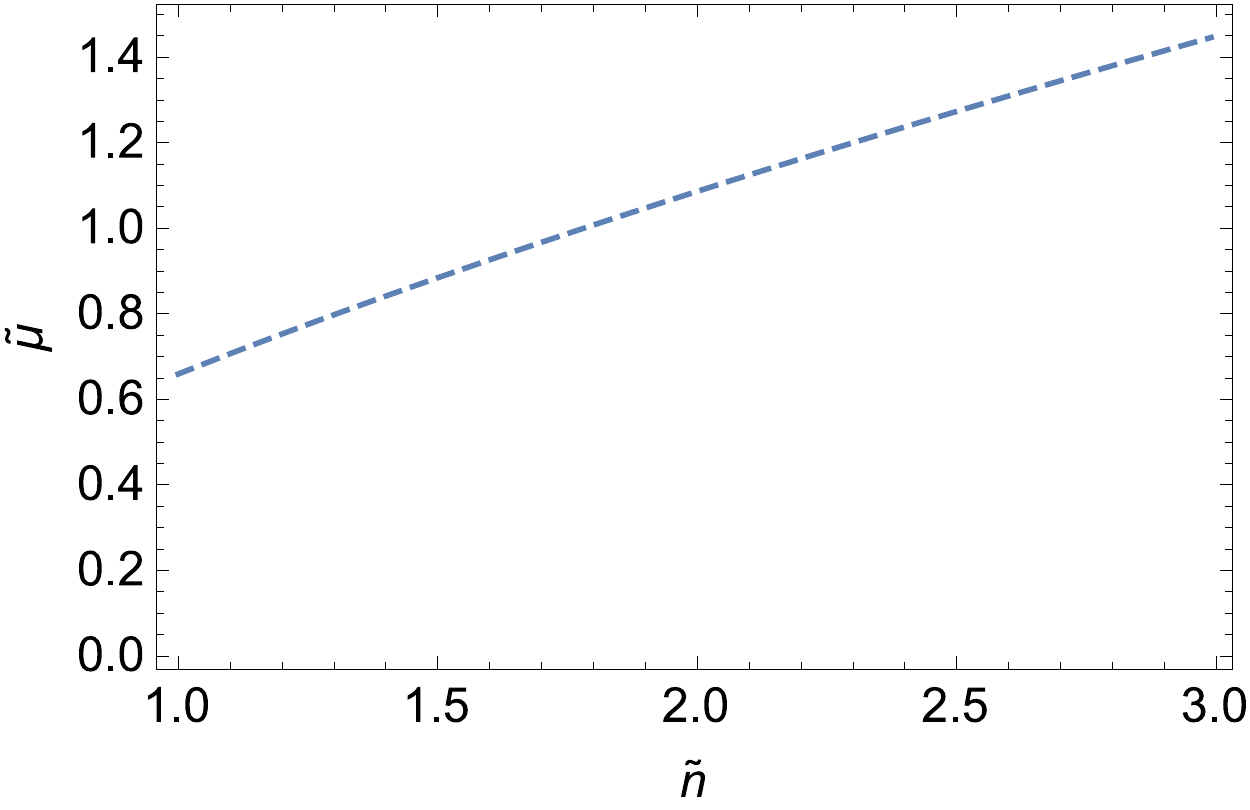}
\caption{Chemical potential $\tilde{\mu}$ of the three-dimensional 
($D=3$) fermionic gas as a function of the average total number density 
$\tilde{n}$. Interaction strength 
$\tilde{g} = 0.1$ and Rabi frequency $\tilde{\Omega} = 0.2$.}
\label{plotmu}
\end{figure}

\begin{figure}[ht]
\includegraphics[width=0.47\textwidth]{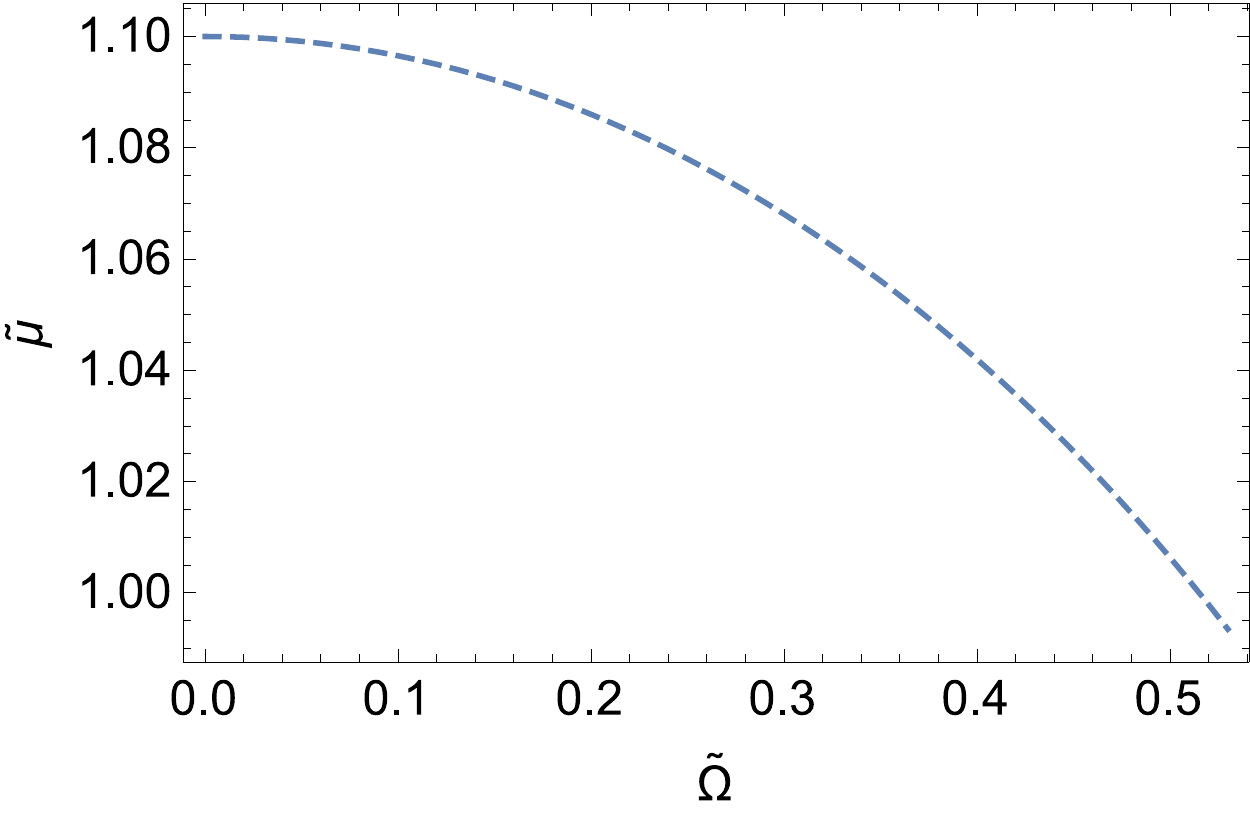}
\caption{Chemical potential $\tilde{\mu}$  of the three-dimensional 
($D=3$) fermionic gas as a function of the Rabi frequency  
$\tilde{\Omega}$.  Interaction strength $\tilde{g} = 0.1$, 
average total number density $\tilde{n} = 2$. }
\label{plotmu2}
\end{figure}

If instead $D=1,3$, the same equations can be solved numerically. 
Clearly, in this case it is much better to work with 
the rescaled equations (\ref{eqmuintsbil2}). 
For $D =3$, the dependence of $\tilde{\mu}$ on $\tilde{n}$ 
at fixed $\tilde{\Omega} =0.2$  and $\tilde{g} = 0.1$ is reported 
in Fig. \ref{plotmu}, while the dependence on $\tilde{\Omega}$ 
at fixed $\tilde{n} =2$ and $\tilde{g} = 0.1$ is shown in Fig. \ref{plotmu2}.

\begin{figure}[ht]
\includegraphics[width=0.46\textwidth]{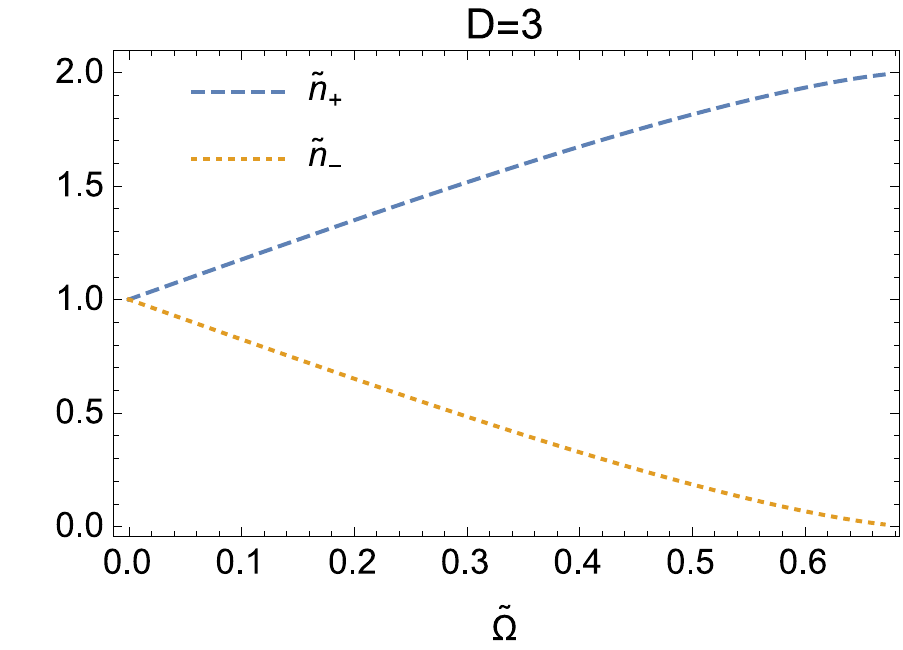}
\caption{Average number densities $\tilde{n}_{\pm}$ 
of the three-dimensional ($D=3$) fermionic gas as a function 
of the Rabi frequency $\tilde{\Omega}$.  
We set in the regime with two bands and we also assumed 
$\tilde{g} = 0.1$, and $\tilde{n} = 2$.}
\label{plotx}
\end{figure}

Numerical solutions  for $\tilde{n}_{\pm}$ are shown instead in 
Fig. \ref{plotx}. We see that as $\Omega$ increases, the same trend 
occurs for $\tilde{n}_+$, while  $\tilde{n}_-$ decreases (so that the total 
rescaled density $\tilde{n}$ stays constant.)

Finally, beyond a critical value for $\tilde{\Omega}$ only the band 
$\lambda_+ (\b{k})$ is populated ($\tilde{n}_- = 0$). Notice also that 
the regime with only a band is achieved at a lower critical $\tilde{\Omega}$ 
when the dimension of the system increases.

\section{Collective dynamics}

The equation of state for the Rabi-coupled Fermi gas determined 
in the last Section allows to turn on the investigation 
of some non-equilibrium properties. Along this line, on the next Sections 
we will study the first sound and the zero sound. 
We focus mainly in the regime where 
both the branches $s = \pm$ are macroscopically populated, so that speaking 
about collective spatial oscillations in their densities makes sense. 
The density perturbation 
giving rise to the sounds can ben created  by a suitable  
blu-detuned laser applied on the gas, producing an effective local repulsive 
potential and a gaussian density hole in the atomic cloud 
(see \cite{sound1,sound2} and citing articles)  \cite{notapert}.    

Given an interacting system with collisional time $\tau$, 
a sound mode of frequency $\omega$ is in the so-called 
collisionless regime if $\tau \, \omega \gg1$. In this case  
the mode is called zero sound \cite{landaustat} and the collisionless 
dynamics is accurately described by the Landau-Vlasov 
equation \cite{landaustat}. Notice that in the three-dimensional 
case the collisional time $\tau$ at $T = 0$ \cite{notetemp} scales as 
\beq
\tau \sim \frac{1}{a^2 \, \sqrt{v_F^{(+)} \, v_F^{(-)}}  \, \sqrt{n_{+} n_{-}} }  \sim \frac{1}{a^2 \, (n_{+} \, n_{-})^{\frac{2}{3}} }
\label{deftau}
\eeq
where $a$ is the scattering length (proportional to the interaction 
strength $g$) and $v_F^{(\pm)} = \sqrt{\frac{2 \, \mu_I^{(\pm)}}{m}} $ 
are the Fermi velocities. 
Instead, under the condition  $\tau \, \omega \ll 1$, the sound mode 
of frequency $\omega$ is in the collisional regime and 
the mode is called first sound. 
The sound depends here on the collective wave motion of the Fermi 
gas, describable by ordinary Navier-Stokes equations (with suitable quantum 
corrections to be included).

We stress at the end that, for both the sounds described above, 
the simultaneous appearance of the two chemical potentials 
$\mu^{(I)}_{\pm}$ in the equilibrium equations of state (as in Eqs. 
(\ref{eqmuintsbil2})) allows to include a priori density oscillations 
of the two species $\pm$ with arbitrary relative phases.  

\section{First sound} 
\label{first}

\subsection*{Derivation of the first sound velocities}

We derive in this Subsection the expressions for the first sound 
velocity in the two-components Fermi gas.  The knowledge of the
equilibrium thermodynamical quantities is again supposed, in the 
light of the studies performed in the Section \ref{equil}. 

In principle, since two fermionic components are involved,  two 
sounds could be expected, corresponding to local fluctuations 
of the total number density $n = n_+ + n_-$ 
and of the difference $\delta n  = n_+ - n_-$. However, the 
single particle energies $\lambda_{\pm}(\b{k})$ in Eq. \eqref{spec} 
differ by an amount $2 \, \Omega$ and the sound related with the
fluctuations of $\delta n$ is gapped. Consequently, an initial very small 
perturbation does not excite the gapped mode. 
For this reason, we concentrate below on the fluctuations of $n$.

As claimed in the previous Section, an hydrodynamic approach,
based on the Navier-Stokes equations, is appropriate in the first sound regime. 
Here we use the equations of rotational hydrodynamics 
\cite{stringari2003} for the local total 
particle densities $n({\bf r},t)$ and local velocity
${\bf v}({\bf r},t)$. These equations read
\beq
\frac{\partial}{\partial t}  n (\b{r},t)
+ {\boldsymbol \nabla} \cdot \big( n (\b{r},t)\, {\bf v} (\b{r},t) \big) = 0  
\label{eqdens1}
\eeq
and
\beqa 
m \frac{\partial}{\partial t} {\bf v} (\b{r},t)
&+& {\boldsymbol \nabla} \Big( \frac{1}{2} m \, |{\bf v} (\b{r},t)|^2 
+ \mu(n (\b{r},t)) \Big) 
\nonumber 
\\
&=& m \, {\bf v} (\b{r},t) \wedge \big( {\boldsymbol \nabla } 
\wedge {\bf v} (\b{r},t) \big) \; , 
\label{eqen}
\eeqa
where $\mu(n ({\bf r},t)) $ 
is the chemical potential of the bulk system, defined in the 
Hamiltonian in Eq. \eqref{H_cont}.\\
Notice that on the right side of 
the equality in Eq. (\ref{eqen}) a rotational term appears.
This term makes the difference between normal collisional 
hydrodynamics and superfluid (irrotational) hydrodynamics 
\cite{stringari2003}, since in the latter case ${\boldsymbol \nabla} 
\wedge {\bf v} ({\bf r},t) = {\bf 0}$. 
However, even for rotational fluids,  the rotational term, quadratic 
in the velocity field, can be neglected 
in the framework of linear-response theory, valid for small perturbations 
of the equilibrium configuration, as the ones considered here.\\
In the following we will work under this assumption, neglecting the 
rotational term and  performing the expansions:
\beq
\left\{ \begin{array}{rl}
n (\b{r},t) = n + \delta n (\b{r},t)\\
{}\\
v (\b{r},t) = 0 + \delta v (\b{r},t)  \, ,
\end{array}  \right.
\label{linsyst}
\eeq
where the symbols $n$ denote the constant equilibrium density defined in 
Section \ref{model}, moreover 
$\delta n(\b{r},t)$ and $\delta v(\b{r},t)$ represents small variations 
with respect 
to the equilibrium configuration.
Similarly, the chemical potential can be expanded around the equilibrium v
alue $\mu = \mu (n)$ as:
\beq
\mu (n (\b{r},t)) \simeq \mu (n)
+\frac{\partial \mu}{\partial n} (n) \, \delta n (\b{r},t) \, .
\label{mulin}
\eeq
We insert now in Eq. (\ref{cons}) the linearizations in Eqs. (\ref{linsyst})  
and (\ref{mulin}), then we derive in $t$ the first obtained equation and 
the second one in $\b{r}$. Finally  we sum each other the two so-obtained 
expressions, arriving to the equation:
\beq
\frac{\partial^2 \delta n (\b{r},t) }{\partial{t^2}} +  \frac{n}{m} 
\frac{\partial \mu (\b{r},t)}{\partial n} \, \nabla^2 \delta n (\b{r},t) 
= 0 \, .
\eeq
The same equation can be solved imposing 
\beq
\delta n (\b{r},t) = C \, e^{i ({\bf k} \cdot \b{r} - \omega t)} \, ,
\eeq 
and solving the consequent algebraic equation for $\omega$. The final result 
for the first sound velocity is (see e.g. \cite{lipparini}):
\beq
v_1 = \frac{\omega}{k} = \sqrt{\frac{n}{m} \, \frac{\partial \mu (n)}
{\partial n}} \, .
\label{formulefirst}
\eeq
In the $D = 2$ case, Eq. \eqref{mu2d} yields:
\beq
v_1^{(2D)} =  \sqrt{\frac{n}{2 \, m \, A_2} \, (1+ g \, A_2)} \, .
\eeq
Notice that the latter result is not valid limit $g \to 0$, where the zero 
sound equations hold instead. 

\begin{figure}[ht]
\includegraphics[width=0.46\textwidth]{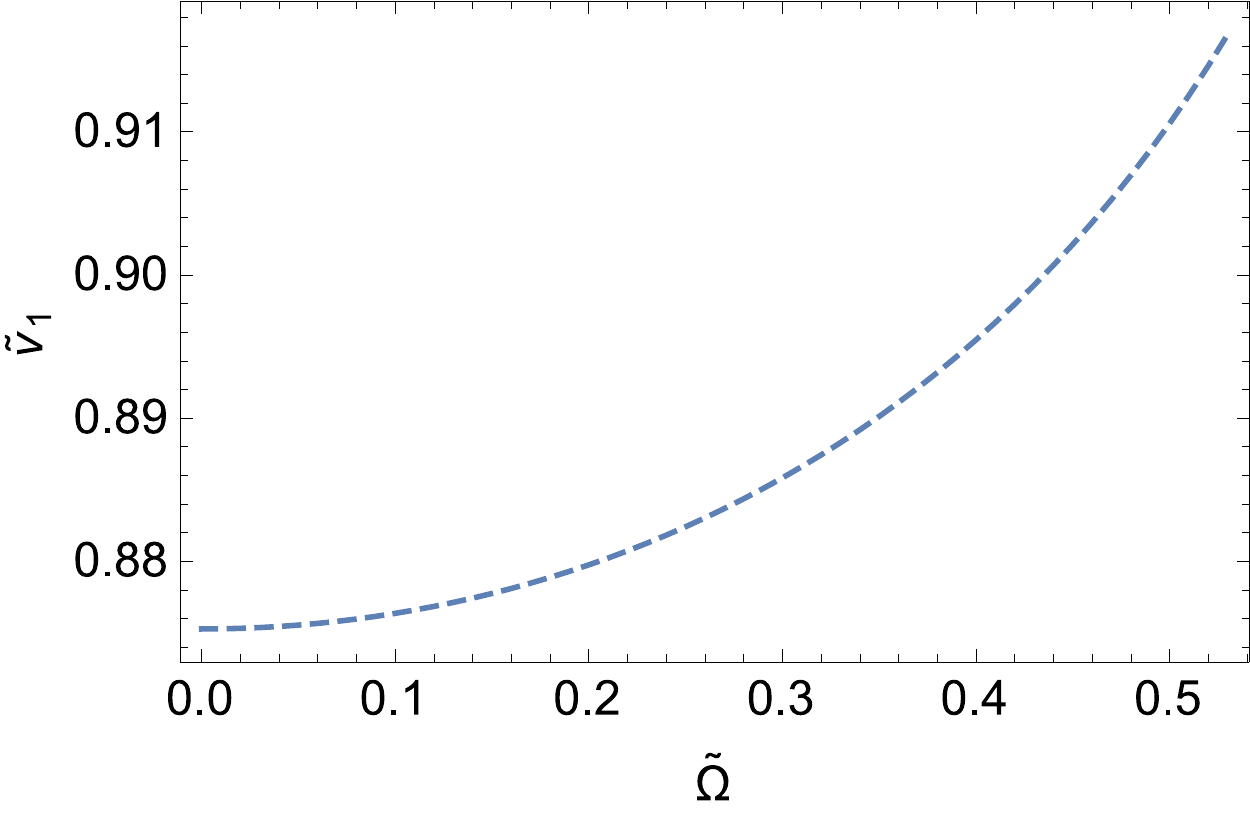}
\caption{First sound velocities $\tilde{v}_1$ as a function of the 
Rabi frequency $\tilde{\Omega}$ for the Fermi gas with $D=3$. 
We assumed interaction strength $\tilde{g} = 0.1$ and 
average number density  $\tilde{n} = 2$.} 
\label{plotv1}
\end{figure}

For the case $D=3$, the explicit dependence of the first sound (rescaled in 
unity of $\sqrt{\frac{\Lambda_D}{m}}$)  $\tilde{v}_1$ on  $\tilde{\Omega}$ at 
fixed $\tilde{n} = 2$ and $\tilde{g} = 0.1$ is shown in Fig. \ref{plotv1}. 
We find an increase of $\tilde{v}_1$ with $\tilde{\Omega}$. 
By increasing $\tilde{\Omega}$ one has more and more atoms in only the state
$+$. At some point the ground state is formed only by atoms in 
this $+$ state and, as a consequence, there is no interaction 
between fermions and the first sound does not exist anymore. 
Thus, for large $\tilde{\Omega}$ the sound  can be collisional, and 
described by Fig. \ref{plotv1}, only for very small momenta. 

\begin{figure}[ht]
\includegraphics[width=0.46\textwidth]{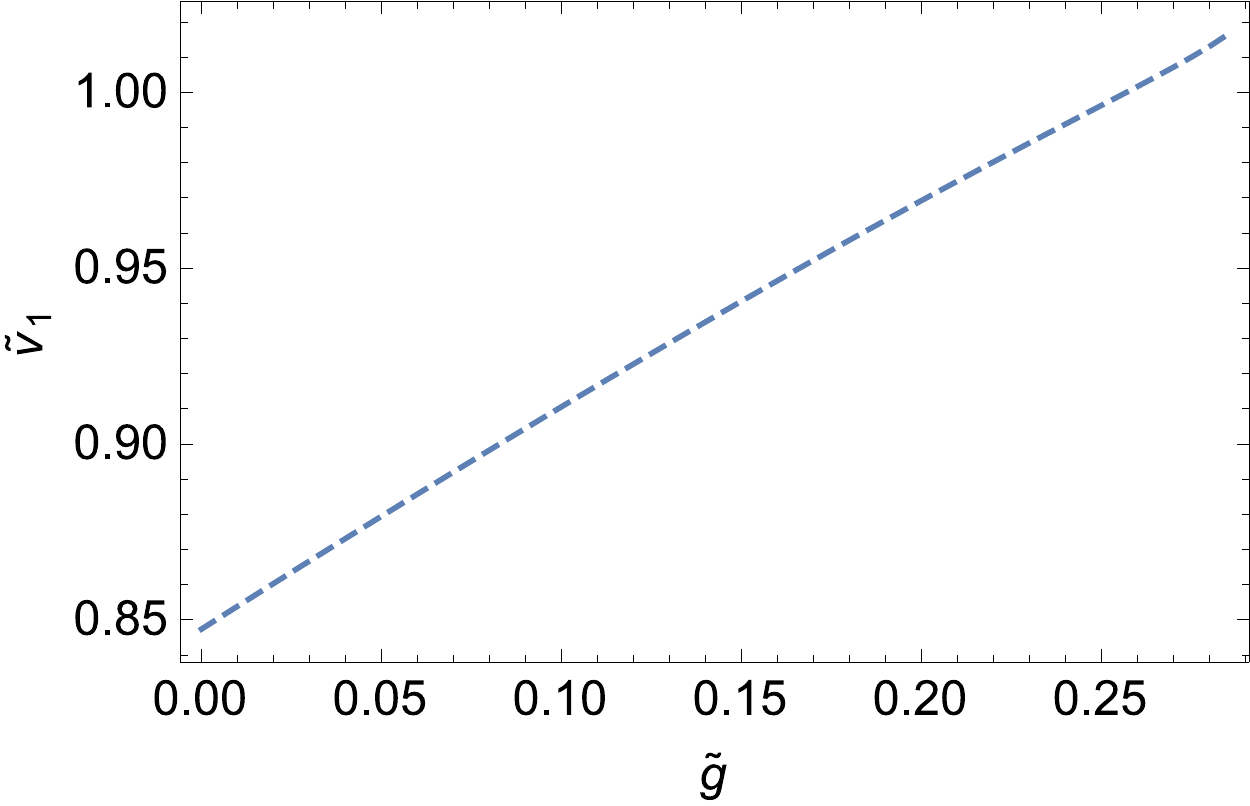}
\caption{First sound velocities $\tilde{v}_1$ as a function 
of the repulsive interaction strength $\tilde{g}$ for the 
Fermi gas with $D=3$. We assumed the Rabi frequency 
$\tilde{\Omega} = 0.5$ and the average number density $\tilde{n} = 2$.} 
\label{plotv1g}
\end{figure}

We also show in Fig. \ref{plotv1g} the dependence of $\tilde{v}_1$ on  
$\tilde{g}$ at fixed $\tilde{n}=2$ and $\tilde{\Omega}= 0.5$, finding 
the opposite behaviour of $\tilde{v}_1$.
 Notice that in both of the plots, regimes where $n_{\pm} \neq 0$
are considered.

We comment finally that in Eq. \eqref{eqen}  we neglected viscosity terms 
$\eta \nabla^2 \b{v}$, the same will be done in the following. Strictly 
speaking, this is possible  with good accuracy close to unitarity 
\cite{turlapov,cao1,cao2}. On the contrary, far from unitary this term leads 
generally to a damping of the oscillations proportional to the factor 
$e^{-\eta \, |\b{r}|}$, nevertheless not affecting the first sound velocity. 

\section{zero sound}

We study in this Section the zero sound in the cases $D= 2$ and $D=3$. 
If instead $D=1$ the Fermi liquid is sustained mostly if $g = 0$, when 
the zero sound coincides with the Fermi velocity $v_F^{(\pm)} = 
\sqrt{\frac{2 \mu_{\pm}}{m}}$.

\subsection*{D=3 case}

The zero sound can be derived exploiting the Boltzmann equation in the 
collisionless regime, named Landau-Vlasov equation \cite{landaustat}:
\beq
\Bigg( \frac{\partial }{\partial t} + \frac{{\bf p}}{m} \cdot 
\b{\nabla}_{\b{r}} 
- \b{\nabla}_{\b{r}}U_{\pm}(\b{r}) \cdot \b{\nabla}_{\b{p}}\Bigg) f_{\pm} 
(\b{r}, {\bf p},t)  = 0 \, .
\label{BV}
\eeq
In this equation $ f_{\pm} (\b{r}, {\bf p},t)$ is the phase-space distribution 
of the fermionic quasiparticles  $\pm$ (assumed both occupied) with position 
${\bf r}$ and  momentum $\b{p}$, having the property 
\beq
\int \mathrm{d} \b{p} \, f_{\pm} (\b{r}, \b{p},t)  = n_{\pm} (\b{r},t) 
 \eeq
 and similarly for the same integral in ${\bf r}$.\\
 The quantities $U_{\pm}(\b{r})$ in Eq. \eqref{BV} are generally the 
potentials acting on the components $\pm$; when the specific interaction 
in Eq. (\ref{H_cont}) is assumed, then $U_{\pm}(\b{r}) = g \, n_{\mp}(\b{r}) $  
are the Hartree potentials introduced in Section \ref{freeint}. 

\begin{figure}[ht]
\includegraphics[width=0.46\textwidth]{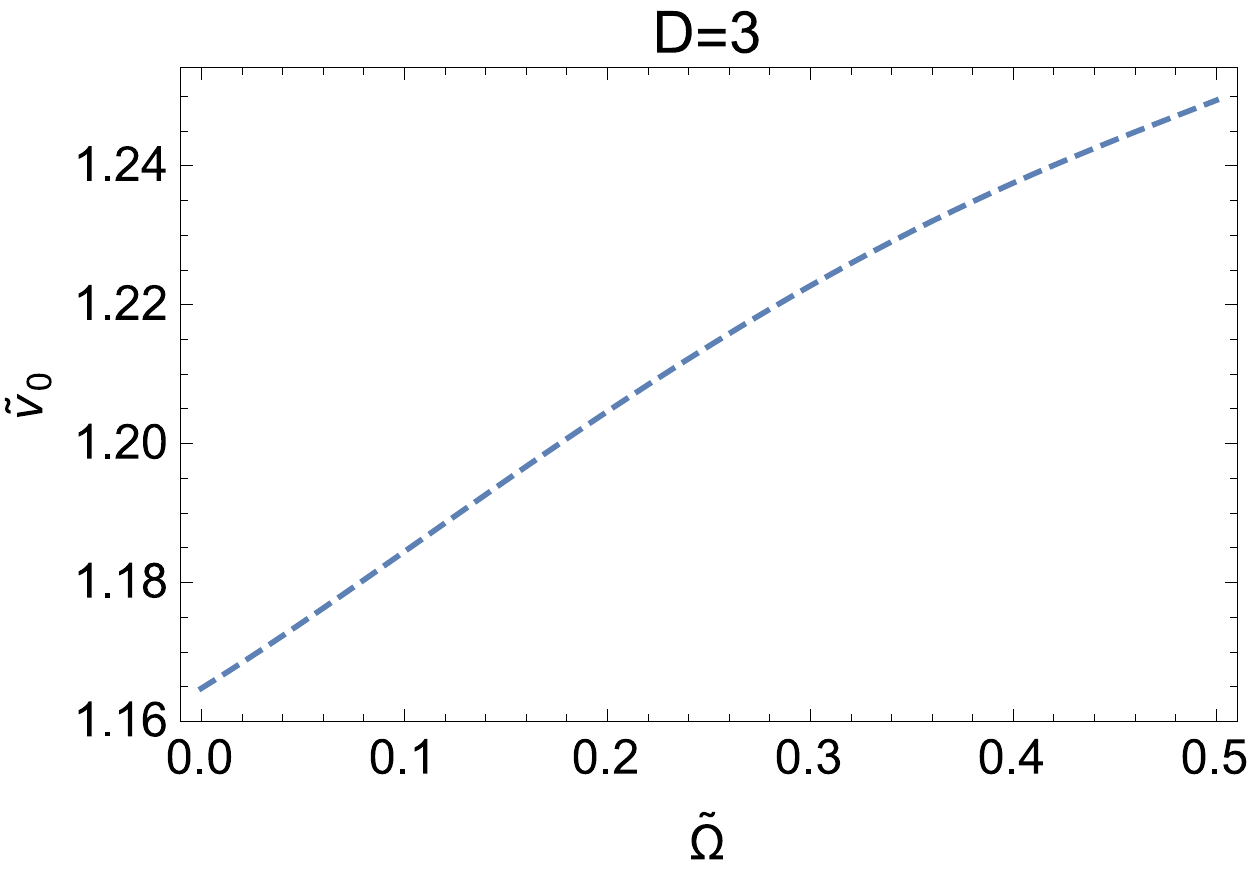}
\caption{Zero sound velocities $\tilde{v} = r_+ \, 
\sqrt{\tilde{\mu}^{(I)}_{+}}$ as a function of the Rabi frequency 
$\tilde{\Omega}$ for the Fermi gas with $D=3$. We assumed 
$\tilde{g} = 0.2$, $\tilde{n} = 2$, and $F_0^{(-)} =1$ 
($m$ chosen accordingly).} 
\label{plotv0}
\end{figure}

\begin{figure}[ht]
\includegraphics[width=0.46\textwidth]{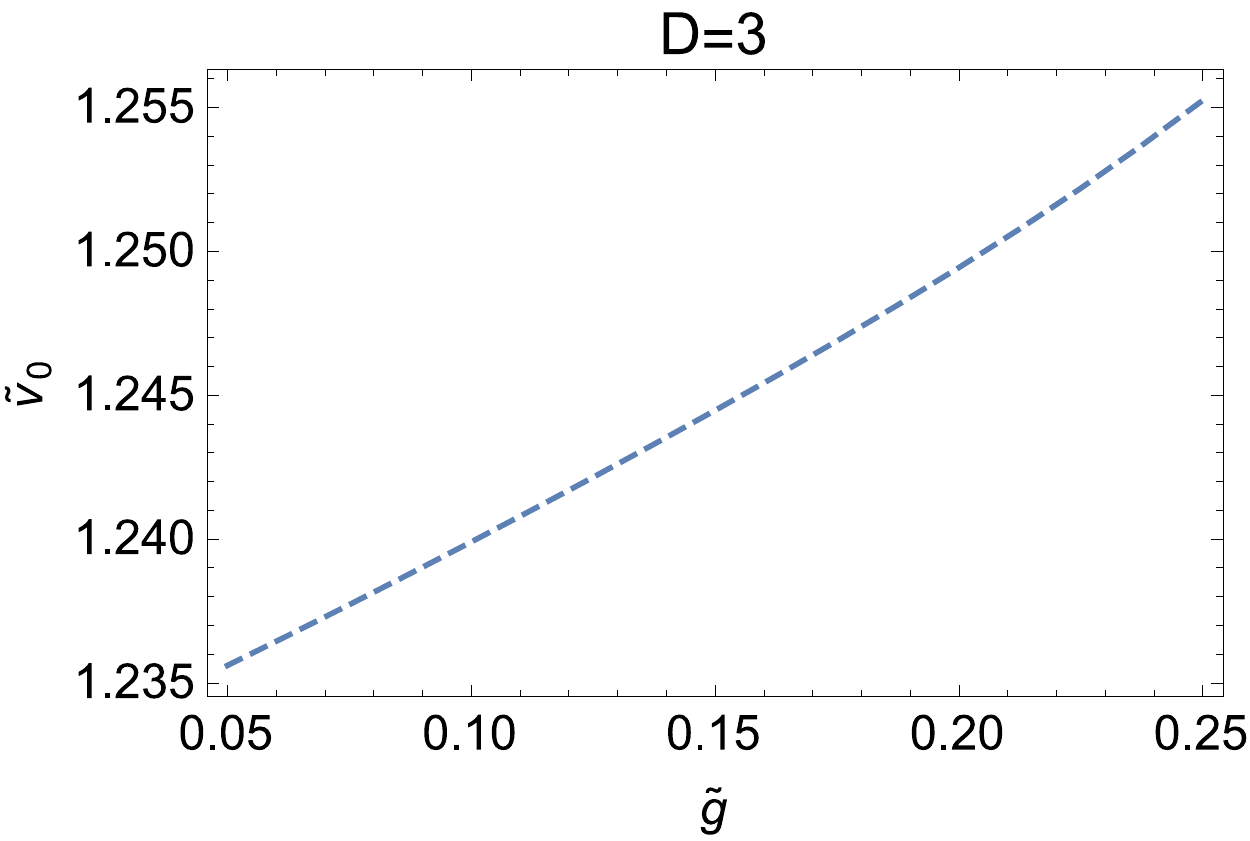}
\caption{Zero sound velocities $\tilde{v} = r_+ \, 
\sqrt{\tilde{\mu}^{(I)}_{+}}$ vs $\tilde{g}$. 
We also assumed $\tilde{\Omega} = 0.5$, $\tilde{n} = 2$,
$D=3$, and $F_0^{(-)} =1$.} 
\label{plotv0g}
\end{figure}

Expanding $ f_{\pm} (\b{r}, {\bf p},t)$ and $n_{\pm}(\b{r})$ around their 
equilibrium values for general $U_{\pm}(\b{r})$ 
\beq
\left\{ \begin{array}{c}
 n_{s} (\b{r},t) = n_s + \delta n_s (\b{r},t)  \\
 {}\\
  f_{s} (\b{r},\b{p},t) = f_{s  ,  \mathrm{eq}}(\b{p}) + \delta f_{s} (\b{r} , 
\b{p} , t) \, ,
\end{array}  \right. \, 
\label{cons}
\eeq
with $s = \pm$, $f_{s  ,  \mathrm{eq}}(\b{p}) = \delta \big(\mu^{(I)}_s - 
\epsilon_{s , \mathrm{eq}} ({\bf p})\big) $ and $\epsilon_{\pm , \mathrm{eq}} 
({\bf p})= \frac{p^2}{2 m} + g \, n_{\pm} $. We obtain (after relabelling 
for sake of brevity $\delta f_{s} (\b{r} , \b{p} , t) \equiv f_{s} (\b{r} , 
\b{p} , t)$):
\begin{multline}
\Bigg( \frac{\partial }{\partial t} + \frac{{\bf p}}{m} \cdot 
\b{\nabla}_{\b{r}} \Bigg) f_{\pm} (\b{r}, {\bf p},t) +  \delta 
\big(\mu_{\pm}^{(I)} - \epsilon_{\pm,\mathrm{eq}} (\b{p}) \big) \, 
\frac{\b{p}}{m} \times \\
\times \b{\nabla}_{\b{r}} \,  \int  \mathrm{d} \b{p}^{\prime} \, 
F(\b{p}, \b{p^{\prime}})  \,  f_{\mp} (\b{r}, \b{p^{\prime}},t) = 0 \, .
\label{BV2}
\end{multline}
The quantity $F({\bf p}, {\bf p^{\prime}})$ is the so-called 
interaction function between the components. This function
measures the change of the quasiparticle energies for small deviations 
of the distribution functions $n_{\pm} (\b{p}) =\int \mathrm{d} \b{r} 
\, f_{\pm} (\b{r}, \b{p},t)$ from the step function $\theta \big(|\b{p}|-
p_{\pm}^{(F)} \big)$ \big($p_{\pm}^{(F)} = \sqrt{2 \,  m \, \mu_{\pm}^{(I)}}$ 
being the Fermi momenta) typical of Fermi gases in equilibrium:
\beq
\delta \epsilon_{\pm,\mathrm{eq}} (\b{p}, t)  = \int \mathrm{d} \b{p^{\prime}} 
\, F({\bf p}, {\bf p^{\prime}}) \, f_{\mp} (\b{r}, \b{p^{\prime}},t)
\eeq 
and it vanishes in absence of interaction \cite{landaustat}. 
For most of the practical cases, one can fix $|\b{p}| = |\b{p}^{\prime} |
= p_{\pm}^{(F)}$, so that
the interaction function can be supposed depending only on  $\hat{p} 
\cdot \hat{p^{\prime}}$.  In these pretty general hypothesis,
Eqs. (\ref{BV}) can be simplified by the linearizing ansatz:
\beq
 f_{\pm} (\b{r}, {\bf p},  t) = \delta \big(\mu^{(I)}_{\pm} - 
\epsilon_{\pm,\mathrm{eq}} (\b{p}) \big) \,
 \Phi_{\pm}(\hat{p}) \, e^{i \, (\b{k} \cdot \b{r} -  \omega \,  t)} \, ,
 \label{ansatz}
\eeq
valid under the further hypothesis that the initial perturbation 
giving rise to the zero-sound has a typical energy scale $U$ such 
that $U \ll \mu^{(I)}_+$ and $U \ll \mu^{(I)}_-$. Assuming the Eq. 
\eqref{ansatz}, we obtain:
\begin{multline}
\Bigg(- \omega \, + \,  \frac{ {\bf p}_{\pm}^{(\mathrm{F})}  \cdot \b{k}}{m} 
\Bigg) \Phi_{\pm} \big(\hat{p}_{\pm}^{(\mathrm{F})}\big) + 
\frac{1}{(2 \pi \hbar)^3} \, p_{+}^{(\mathrm{F})} \, p_{-}^{(\mathrm{F})} \, 
\big(\hat{p}_{\pm}^{(\mathrm{F})}  \cdot \b{k} \big) \times\\
\times  \int \,  \Phi_{\mp}  \big(\hat{p^{\prime}} \big) \, 
F\big(\hat{p}^{(\mathrm{F})}_{\pm} \cdot \hat{p^{\prime}}\big) \,
 \mathrm{d} \hat{p^{\prime}} = 0 \, .
\label{zerofinal} 
\end{multline}
In the absence of interaction, $F\big(\hat{p}^{(\mathrm{F})}_{\pm} \cdot 
\hat{p^{\prime}}\big) = 0$, Eqs. \eqref{zerofinal} yield correctly two 
Fermi velocities: 
\beq
\frac{\omega}{|{\bf{k}}|}  \equiv v =  v_{\pm}^{(F)} = \sqrt{\frac{2 \, 
\tilde{\mu}^{(I)}_{\pm}}{m}} = \frac{p_{\pm}^{(\mathrm{F})}}{m} \, .
\eeq
Assuming now the point-like repulsion in Eq. (\ref{H_cont}), so that in 
Eqs. (\ref{BV}) it results $U_{\pm}(\b{r}) = g \, n_{\pm}(\b{r})$, the 
interaction function turns out to be  constant and equal to  $2 \, g$,  and 
the latter equation reduces to
\begin{multline}
\big(-  r_{\pm} +  \mathrm{cos} \, \theta \big) \, 
\Phi_{\pm} \big(\hat{p}_{\pm}^{(\mathrm{F})}\big) + 
\mathrm{cos} \, \theta \, F_0^{(\mp)} \times \\
\times \int  \frac{ \mathrm{d} \hat{p^{\prime}} }{4 \pi} \, \, \Phi_{\mp}  
\big(\hat{p^{\prime}} \big) = 0  \, ,
\label{pp}
\end{multline}
with the adimensional quantities
\beq
F_0^{(\pm)} \equiv \frac{g \, p_{\pm}^{(F)} \, m}{\pi \hbar^3} 
= \left(\frac{6}{\pi}\right)^{\frac{1}{3}} \, \frac{g \, m}{\hbar^2} \, 
n_{\pm}^{\frac{1}{3}} 
\eeq
and
\beq
  r_{\pm} = \frac{\omega}{|\b{k}|  v_{\pm}^{(F)}} = \frac{v}{ v_{\pm}^{(F)}} \, .
\eeq
Eqs. (\ref{pp}) can be also derived by a random phase approximation (RPA) 
approach for unbalanced Fermi gases, used e.g. in \cite{stringari2012}. 
Moreover they can be solved directly by the substitution \cite{landaustat}
\beq
\Phi_{\pm}  \big(\beta\big) = C_{\pm} \, \frac{\mathrm{cos} \, 
\beta}{r_{\pm} - \mathrm{cos} \, \beta} \, ,
\label{sub}
\eeq
leading finally to the system \cite{notadef}
\beq
\left\{ \begin{array}{rl}
\Gamma =  F_0^{(-)} \, \Big(-2 + r_- \, \mathrm{ln} \, 
\frac{r_- +1}{r_- -1}\Big) \\
{}\\
\frac{1}{\Gamma} =  F_0^{(+)} \, \Big(-2 + r_+ \, \mathrm{ln} \, 
\frac{r_+ +1}{r_+ -1}\Big)  \, ,
\end{array}  \right.
\label{lin}
\eeq
where $\Gamma = \frac{C_+}{C_-}$. We also define: 
\beq
 a  \equiv \frac{F_0^{(+)}}{F_0^{(-)}} = \frac{v^{(F)}_+}{v^{(F)}_-}  = 
\sqrt{\frac{\tilde{\mu}_+^{(I)}}{\tilde{\mu}_-^{(I)}}} \, .
 \label{defa}
\eeq
In this way, the condition 
\beq
v = v^{(F)}_+ \, r_+ = v^{(F)}_- \, r_-
\label{vela2}
\eeq
(we look for a unique sound, corresponding with the situation in 
the collisional regime) results into the other one:
\beq
r_- = a \, r_+ \, .
\eeq
We insert now the last in relation in Eq. \eqref{lin} and we solve 
the so-obtained system in $r_+$.
In this way, exploiting Eq. \eqref{vela2}, we arrive finally to the  
zero sound velocity (rescaled in unity of 
$\sqrt{\frac{2 \, \Lambda_D}{m}}$ \Big) 
\beq
\tilde{v} = r_+ \, \sqrt{\tilde{\mu}^{(I)}_{+}} \, .
\eeq
The results are reported  in Fig. \ref{plotv0} for $\tilde{\Omega}$ varying 
and fixed $\tilde{n}=2$ and $\tilde{g} =0.2$ 
and  in Fig. \ref{plotv0g} for $\tilde{g}$ varying and fixed $\tilde{n} =2$ 
and $\tilde{\Omega} = 0.5$. Again in both of the plots, regimes where 
$n_{\pm} \neq 0$ are considered. 

\subsection*{D = 2 case}

In this case Eq. (\ref{pp}) becomes:
\begin{multline}
\big(- r_{\pm} +  \mathrm{cos} \, \theta \big) \, \Phi_{\pm} 
\big(\hat{p_{\pm}}^{(F)} \big) +  
 \mathrm{cos} \, \theta \, F_0^{(\mp)} \times\\
\times \int  \frac{\mathrm{d} \hat{p^{\prime}}}{2 \pi} \, \, 
\Phi_{\mp}  \big(\hat{p^{\prime}} \big) = 0  \, ,
\label{pp2d}
\end{multline}
and the substitution in Eq. (\ref{sub}) leads to the system:
\beq
\left\{ \begin{array}{rl}
\Gamma = F_0^{(-)} \, \Big( -1 + \frac{r_-}{\sqrt{r_-^2 -1}}\Big) \\
{}\\
\frac{1}{\Gamma} = F_0^{(+)} \, \Big( -1 + \frac{r_+}{\sqrt{r_+^2 -1}}\Big)  \, .
\end{array}  \right.
\label{lin2d}
\eeq 
Notice that the differences between the systems in Eq. ({\ref{lin2d}}) and in 
Eq. (\ref{lin}) are due to the different angular integrations  in Eq. 
(\ref{pp}) and in Eq. (\ref{pp2d}). The system in Eq. (\ref{lin2d}) 
can be solved numerically in $r_+ $, as for the $D=3$ case. 
The behaviours for the final rescaled velocity 
$\tilde{v} = r_+ \, \sqrt{\tilde{\mu}^{(I)}_{+}}$ are very qualitatively similar 
to the $D=3$ case, reported  in Figs. \ref{plotv0} and \ref{plotv0g}. 
We notice finally that for the present case, $D =2$, a study of the zero 
sound has been performed by a RPA approach in \cite{stringari2012} 
and \cite{sala-wr,toigo2015} also in the presence of a Rashba coupling. 

\section{Conclusions}

In this paper we have analyzed the equation of state 
of a two-component repulsive Fermi gas under the application of a 
Rabi coupling. As main application we have investigated 
the behavior of first sound and zero sound 
after a local perturbation of the uniform density.  
Notably the application of the Rabi coupling appears 
as an effective  experimental strategy to tune the sounds by hand, 
varying the Rabi frequency $\Omega$. 
Finally, in the Appendix some density profile in presence 
of an harmonic external trap, 
mostly used in current experiments, are derived in local density 
approximation, as well as the densities and the chemical potentials 
at the center of the trap, where the measurements for the sounds are usually 
performed. Our findings can be relevant for current experiments 
in ultracold gases, where the application of Rabi couplings is widely 
used for various purposes, as mentioned in the Introduction.

\section*{Acknowledgments}

The authors thank Roberto Onofrio, Flavio Toigo, and Andrea Trombettoni
for useful discussions and acknowledge the University of Padova. 
LS acknowledges for partial support the 2016 BIRD project 
"Superfluid properties of Fermi gases in optical potentials" of the 
University of Padova. 

\section*{Appendix: Trap effects}

In all the examples considered so far we analyzed uniform gases in a 
continuous $D$-dimensional space. At variance, in many other real experiments 
the space-density is made non uniform by the presence of a confining harmonic 
potential. In these conditions, the sound velocity is generally 
measured in the centre of the trap, so that the results of the previous 
Sections hold using the 
atomic densities and chemical potentials in this point. Although these 
quantities can be as well measured directly in the centre of the trap, 
it is interesting to derive analitically some density profile in the 
presence of an harmonic trap; this is the aim of this Section. \\   
 The trapping harmonic  potential has the general form:
 \beq
 U(\b{r}) = \frac{1}{2} \, m \, (\omega_x^2 \, x^2 +\omega_y^2 \, 
y^2 + \omega_z^2 \, z^2) \, .
 \eeq
 The effect on the trap can be taken into account in local-density 
approximation, defining two space dependent 
 chemical potentials as \cite{capuzzi2006}:
 \beq
 \mu^{(I)}_{\pm} (\b{r}) = \mu^{(I)}_{\pm} - U(\b{r}) \, ,
 \eeq 
 being $\mu^{(I)}_{\pm}$
 the equilibrium chemical potentials (defined as in Eq. \eqref{shift}\big) 
at the center of the trap for the components $\eta_{\pm}$, and  
two  space-dependent densities $n_{\pm} ({\bf r})$ related to 
$\mu^{(I)}_{\pm} ({\bf r})$ as in the previous Section: $\mu^{(I)}_{\pm}(\b{r})
 = \Big(\frac{n_{\pm}(\b{r})}{A_D}\Big)^{\frac{2}{D}}$.
 The quantities $\mu_{\pm}$ can be found by solving the equation:
 \beq
 \int_{V_+} \, \mathrm{d} \b{r} \,  n_{+}(\b{r}) + \int_{V_-} \, 
\mathrm{d} \b{r} \,  n_{-}(\b{r}) = N \, ,
 \label{const}
 \eeq 
 being $N$ the total number of loaded atoms, assumed known,  and  
$V_{\pm}$ are the volumes of the trap, whose extensions, characterized 
by the radii $R^{(\pm)}_F = \sqrt{\frac{2 \, \mu^{(I)}_{\pm}}{m \, \omega^2}}$,  
are limited by the condition $\mu^{(I)}_{\pm} (\b{r}) >0$. 
 
The total density $n (\b{r})$ is then $n (\b{r})= n_+ (\b{r}) + n_- (\b{r)}$, 
with $n_{\pm} (\b{r})$ related to $\mu_{\pm}^{(I)} (\b{r})$ as above.

We assume first the non interacting case $g = 0$ (where $\mu^{(I)}_{\pm} 
= \mu_{\pm} = \mu \pm \hbar \Omega$). 
If $D=1$ ($\omega_y \, , \, \omega_z \gg \omega_x \equiv \omega$ 
and $\hbar \omega_y \, , \, \hbar \omega_z \gg \mu $) the 
integration of $n_{\pm} (\b{r})$ up to $R^{(\pm)}_F$ yields:
\beq
\mu = \frac{3}{8 \sqrt{2}} \, \frac{N}{A_1} \, m^{\frac{1}{2}} \omega \, .
\label{1dcase}
\eeq
If $D=2$ ($\omega_z \gg \omega_x =   \omega_y \equiv \omega$ and 
$\hbar \omega_z \gg \mu$), similar calculations lead to:
\beq
\mu =  \sqrt{\frac{N}{2 \pi A_2} m \, \omega^2- (\hbar \Omega)^2}  \, .
\label{2dcase}
\eeq
Notice that in this case the positiveness of the argument in the square 
root in Eq. (\ref{2dcase}) is always fulfilled when the two branches $\pm$ 
in Eq. (\ref{spec}) are both populated. 

In the case $D=3$, $\mu$ can be obtained as the unique real solution 
of the third order algebraic equation
\beq
\mu^3+ 3 \, (\hbar \Omega)^2 \, \mu = \frac{15}{32 \pi \sqrt{2}} \, 
\frac{N \, m^{\frac{3}{2}} \, \omega^3}{A_3} \, ,
\eeq  
where $\omega_x = \omega_y = \omega_z \equiv \omega$. 
Although the analitical expression of this solution is not particularly 
enlightening, its qualitative analysis indicates
a growth for  $\mu$ as  $\omega$ increases, and the opposite trend when 
$\Omega$ is varied. Both these behaviours are the ones 
expected from intuition and are also found for the cases $D=1,2$, Eqs. 
(\ref{1dcase}) and (\ref{2dcase}). 
Remarkably, if $D=1$ no dependence on $\Omega$ is found: 
this effect has the same origin of the homogeneous 2D case and it is  
simply due to a constant density of states. 
Notice that for every $\omega >0$ and $\Omega >0$, 
it always exists a solution $\mu >0$. 

In the interacting case $g \neq 0$, $\mu$ and $\mu^{(I)}_{\pm}$ 
can be found from the equations:
\beq
\Bigg\{ 
\begin{array}{c}
n_+ (\b{r}) = A_D \, \big(\mu + \hbar \Omega - g \, n_-({\bf r})- 
\frac{1}{2} m \, \omega^2 |{\bf r}|^2\big)^{\frac{D}{2}}\\
{}\\
n_- (\b{r}) = A_D \, \big(\mu - \hbar \Omega  -g \, n_+({\bf r})- 
\frac{1}{2} m \, \omega^2 |{\bf r}|^2\big)^{\frac{D}{2}}  \, .
\end{array}
\label{syst}
\eeq
A semi-analitical solution of the system in Eq. (\ref{syst}) is  available 
in general, we give here details about the case $D = 2$. Under this condition 
we obtain  $\big( g \, A_2 \ll 1 \big)$:
\beq
n_{\pm} (\b{r}) = A_2 \Bigg(\frac{\mu}{1+g \, A_2} \pm \frac{\hbar \Omega}
{1- g \, A_2}- \frac{1}{2 \, (1 + g \, A_2)} \, m \, 
\omega^2 |{\bf r}|^2 \Bigg) \, . 
\label{densexpr}
\eeq
In this way, the condition $\mu^{(I)}_{\pm} (\b{r}) = 0$ (equivalently 
$n_{\pm} (\b{r}) = 0$) gives:
\beq
R_F^{(\pm)} (\mu) = \sqrt{\frac{2 \, (1 + g \, A_2)}{m \, \omega^2} 
\, \Bigg(\frac{\mu}{1 + g\, A_2} \pm \frac{\hbar \, \Omega}{1- g \, A_2} 
\Bigg)} \, .
\eeq
As  functions of $R_F^{(\pm)}(\mu)$, exploiting Eqs. \eqref{const} 
and \eqref{densexpr} we obtain finally a second order algebraic equation 
for the chemical potential $\mu \neq 0$ at the center of the trap:
\begin{multline}
\frac{4}{m \, \omega^2} \, \frac{1}{1+g \, A_2} \, 
\mu^2 - \frac{2}{m \, \omega^2} \, \frac{\hbar \, \Omega}
{1 -g \, A_2}  \, \mu  \, +\\
+ \frac{4}{m \, \omega^2} \, \big(\hbar \, \Omega \big)^2 \, 
\frac{1 + g \, A_2}{(1-g \, A_2)^2}  - \frac{N}{\pi \, A_2} = 0 \, .
\label{eqmutrap}
\end{multline}
When $ g \, A_2 \ll 1 $ and $N \gg1$, the last equation has a positive 
solution only. Exploiting it, the density profile $n (\b{r})$ can be found 
by summing Eqs. \eqref{densexpr}.


\begin{thebibliography}{99}
\bibitem{notapert}
Another net effect of the initial perturbation is to create locally 
superpositions  $\alpha \, |+\rangle + \beta \, |- \rangle$, 
$\beta=\sqrt{1-\alpha^2}$, with $\alpha$ randomly distributed in the 
complex domain of radius $1$.  This fact does not spoil the description 
of the (zero and first) sound propagation by the interactions between 
$|\pm\rangle$ packets. Indeed it is straightforward to check that, 
thanks to the Fermi statistic, any state (in particular imagined before 
the interaction) composed by the tensor product of two superpositions 
as above can be rewritten (in second quantization) as $c \, |+-\rangle$.
\bibitem{notetemp}
Notice that the temperature plays the most relevant role for the collisional time in
a Fermi liquid \cite{landaustat,baym}.
\bibitem{notadef}
The right-hand side of Eqs. \eqref{lin} differs from the corresponding 
Eqs. in \cite{landaustat} for a $\frac{1}{2}$ factor. 
This difference depends on the notations adopted here.
To verify the correctness of our expressions,  we checked that 
from them we obtain $a \to 1$ and $\Gamma \to1$ when $\Omega \to 0$ 
(see definitions below Eqs. \eqref{lin}), as expected from physical 
considerations. Vice-versa, the mentioned  $\frac{1}{2}$ factors
do not allow to recover the same limit conditions.
\end{thebibliography}
\end{document}